\documentclass[prb,twocolumn]{revtex4}

\usepackage{amsmath, amsthm, amssymb}

\newcommand{\av}[1]{\ensuremath{\langle{#1}\rangle}}
\newcommand{\rss}{\rm\scriptscriptstyle}

\newcommand{\ds}{\displaystyle}

\usepackage{graphicx}
\usepackage{psfrag}

\begin{document}

\title{Density functional theory of vortex lattice melting in layered
superconductors: \\ a mean-field--substrate approach}
\author{A.\ De\ Col$^{1}$, G.I.\ Menon$^{2}$, and G.\ Blatter$^{1}$}
\affiliation{$^{1}$Theoretische Physik, ETH-Z\"urich, 
CH-8093 Z\"urich, Switzerland}
\affiliation{$^{2}$The Institute of Mathematical Sciences, 
C.I.T. Campus, Taramani, Chennai 600\ 113, India }
\date{\today}


\begin{abstract}

We study the melting of the pancake vortex
lattice in a layered superconductor in the limit of
vanishing Josephson coupling. Our approach combines
the methodology of a recently proposed mean-field
substrate model for such systems with the classical
density functional theory of freezing. We derive a 
free-energy functional in terms
of a scalar order-parameter profile and use it to
derive a simple formula describing the temperature
dependence of the melting field. Our theoretical
predictions are in good agreement with simulation data. 
The theoretical framework proposed is thermodynamically 
consistent and thus capable of describing the negative 
magnetization jump obtained in experiments. Such consistency
is demonstrated by showing the equivalence of our
expression for the density discontinuity at the
transition with the corresponding Clausius-Clapeyron
relation.  

\end{abstract}

\maketitle


\section{Introduction}

The melting of the vortex lattice is one of the
most striking aspects of the phenomenology of
high-temperature superconductors
\cite{giannireview,Brandt1995,vortexchapter}. 
It is now both theoretically
\cite{nelson88,houghton89,brandt89,blatter96,Nordborg1997}
and experimentally\cite{zeldov95,schilling96}
established that enhanced thermal fluctuations
render the Abrikosov flux-line lattice unstable
to a vortex liquid over a large part of the
$B$-$T$ phase diagram in these materials.
Much interest has been devoted to the melting
transition in extremely anisotropic materials
\cite{sengupta91,ryu92,blatter96,dodgson00} such
as Bi$_2$Sr$_2$CaCu$_2$O$_8$.  The large value of the
anisotropy\cite{colson03} in these materials motivates
their modeling in terms of a one dimensional array of 
magnetically-coupled two-dimensional superconducting layers with
no inter-layer Josephson coupling. Vortex lines are
then described by linear arrangements (stacks) of
pancake vortices\cite{clem91}, whose highly anisotropic
interactions are mediated only by the magnetic field.

The first theories for the melting of the vortex 
lattice\cite{houghton89}  were based on the
Lindemann criterion\cite{Lindemann1910}.
The Lindemann melting criterion yields qualitatively
accurate estimates for the location of the first-order
melting transition. However, since it is based on the crystal
properties only, it describes an instability rather than the 
melting itself. More detailed approaches, in particular 
analyses encompassing both solid and liquid phases,
are thus required for a quantitative description of the 
transition and of the discontinuities in
thermodynamic quantities upon melting.

Apart from numerical simulations\cite{fangohr03},
the extreme limit of zero Josephson coupling
between the layers has been analysed
through classical Density Functional
Theories\cite{sengupta91,menon96,cornaglia00}
(DFT) and using a ``mean-field" substrate
approach\cite{dodgson00}.  Even if only the
electromagnetic interaction is retained, the analysis
of the melting transition is a challenging task due
to the long-range character of the interaction.
At large magnetic fields the strong in-plane
vortex repulsion dominates over the out-of-plane
interactions and the melting line approaches the
2D melting temperature $T_\mathrm{m}^{\rss 2D}$ of
the individual planes.  On the other hand, at low
fields, ignoring the possibility of low-field
reentrance\cite{nelson88}, only few
pancake vortex stacks are present in the system.
Thermal fluctuations trigger the evaporation
of isolated vortex stacks\cite{clem91} at $T_{\rss BKT}$, in
correspondence to the zero-field two-dimensional
Berezinskii-Kosterlitz-Thouless (BKT) transition.
For intermediate values of the external magnetic
field the melting line interpolates between the BKT
transition temperature $T_{\rss BKT}\lesssim T_c$
close to the material critical temperature and
the two-dimensional lattice melting temperature
$T_\mathrm{m}^{\rss 2D} \ll T_c$. In this regime,
the full three-dimensional character of the
system must be accounted for.  The weak long-range
out-of-plane interaction motivates a mean-field
treatment\cite{dodgson00}, in which the 2D pancake
vortex system in each layer is considered separately,
with the effects of the other layers included
via a self-consistent substrate potential. Extensive
Monte Carlo and molecular dynamics simulations have
confirmed the validity of this approximation
and the accuracy of its results\cite{fangohr03}.

If the results of previous DFT approaches to
the numerical melting line are compared with the
simulation data,  consistent results are obtained
at magnetic fields larger than about $0.5 B_\lambda$
($B_\lambda = \Phi_0/\lambda^2$, where $\Phi_0=hc/2e$
is the unit flux and $\lambda$ the planar penetration depth).  
However, at lower fields, the disagreement between the 
two theories is substantial, with the DFT melting line shifted
to much higher temperatures in comparison with the
simulation data.  Another difficulty with previous
DFT studies is associated with the prediction of the
sign of the magnetization jump across
the transition.  As it is well known, the vortex lattice
exhibits a {\em negative} jump in magnetization
upon freezing, leading to a solid phase which is
less dense than the liquid, as in the ice-water
transition. Such anomalous behaviour was first
obtained in direct measurements of the magnetization
discontinuity across the transition
\cite{zeldov95}. These observations are
consistent with the negative slope of the melting
line in the $T$-$H$ phase diagram taken together with
the Clausius-Clapeyron relation.  However, earlier
DFT analyses of the freezing transition in the pancake
vortex system obtained thermodynamically inconsistent
results\cite{cornaglia00}, reporting a positive density 
change and a negative slope of the melting line.

In this paper, we adapt the DFT analysis to incorporate
ideas from the substrate approach; our analysis
is simpler and more accurate in comparison with simulation
data. Moreover, it addresses and solves the problems with the thermodynamic consistency faced in previous studies. The approach
of this paper was first described in a recent letter
where the effects of a surface on the melting
transition have been studied\cite{decol05b}. In the present work we concentrate
on the implications of our approach for the bulk
transition in the infinite system.

The main methodological difference between the work
described here and previous DFT approaches lies in the
determination of the direct correlation function. While
in previous work, this correlation function was derived
{\it ab initio} from the microscopic vortex interaction
using the hypernetted chain approach or more elaborate
extensions, here it is obtained by combining results
from Monte Carlo simulations of 2D logarithmically
interacting particles, i.e., the One Component Plasma
(OCP), with the substrate potential approach. The
correlations of the 2D OCP are used to describe the
effects of the strong in-plane  vortex interaction,
while the substrate potential accounts for the
out-of-plane contributions. Within this new approach
we obtain a simple expression for the free-energy
which can be extended to the inhomogeneous case.

We also prove the thermodynamic consistency of our
approach by showing how to obtain the negative density
jump across the transition, fully consistent with the 
Clausius-Clapeyron equation. We do this by including
a constraint which enforces an integer number of
particles per unit cell\cite{laird87}. Contrary to
earlier claims\cite{cornaglia00}, no higher order
correlation functions (three point or more) are
needed to obtain the anomalous sign of the density
discontinuity. 


\section{Model}

Strongly anisotropic layered superconductors, such as the
Bi- based compounds, are conveniently described in terms of the
Lorentz-Doniach model\cite{Lawrence1971}. The basic 
topological objects are two dimensional vortices (pancake 
vortices) with a core limited to a single superconducting layer. 
The interaction between pancake vortices is strongly anisotropic 
due to the underlying layered structure. In Fourier space the potential
reads\cite{giannireview}
\begin{equation}\label{pot}
  V(K,k_z) = \frac{\Phi_0^2 d^2}{4\pi} \frac{K^2 + k_z^2}
  {K^2[1+\lambda^2(K^2+k_z^2)]};
\end{equation}
here $d$ is the layer spacing and $\lambda$ the
bulk penetration depth in the plane.
When placed on the same layer, pancake vortices feel a strong 
repulsive interaction, logarithmically
dependent on their separation, 
\begin{equation}\label{v0}
 V_0({\bf R}) = -2 \varepsilon_0 d \ln (R/\xi),
\end{equation}
where $\varepsilon_0 = (\Phi_0/4\pi\lambda)^2$ is the vortex line energy 
and $\xi$ the (in-plane) correlation length.
The interaction between vortices residing
on different layers is attractive.  Fourier transforming (\ref{pot}) back 
along the $z$ coordinate we obtain
\begin{align}\label{vk}
   V_{z}(K) = & 
  - \frac{2\pi\varepsilon_0d^2}{K^2 \lambda^2 K_+} 
 e^{-K_+|z |},
\end{align}
with $K_+ = \sqrt{1/\lambda^2 + K^2}$. Finally, transforming to 
planar real coordinates as well one obtains the potential 
in real space
\begin{equation}\label{vz}
  V_z({\bf R}) = - \varepsilon_0 d \frac{d}{\lambda^2}
  \int_0^{+\infty} dK\, \frac{J_0(KR) e^{-K_+|z|}}{KK_+}.
\end{equation}
For large in-plane separations $R \gg \lambda$
a logarithmic attractive interaction is obtained
$V_{z \neq 0}(R) \sim - (d/\lambda) e^{-z/\lambda} V_0(R)$, 
suppressed by a factor
$d/\lambda$ when compared with the in-plane one. 
This out-of-plane interaction 
decays exponentially in the $z$ direction 
over a distance $\lambda$,
extending over a large number ($\lambda/d$) of layers.
The strong anisotropy in the vortex interaction allows us to
separate the strong in-plane repulsion
from the weak out-of-plane interaction. The overall effect
of the latter can then be accounted for via an effective substrate 
term\cite{dodgson00}.


\section{Classical Density Functional Theory} 

We follow the classical Density Functional Theory of freezing of
Ramakrishnan and Yussouf described in Refs.\ \onlinecite{ramakrishnan79,chaikinlubensky}
by choosing the uniform liquid as the reference state and estimating 
the difference in free energy due to the appearance of finite density 
modulations. For simplicity, we consider here a generic three-dimensional 
system of interacting particles; the modifications needed to describe 
the strongly anisotropic pancake vortex system are presented in 
the next section.   

The spatial arrangement of particles is described through 
the density field
\begin{equation}
  \rho_\mu ({\bf r})= \sum_{i = 1}^N 
  \delta({\bf r}-{\bf r}_i),
\end{equation}
where ${\bf r}_i$ is the position of the $i$-th particle (the index $\mu$ 
emphasizes that  $\rho_\mu ({\bf r})$ describes the non-averaged 
microscopic density).
To analyze the finite temperature behavior of the system
we consider the density $\rho({\bf r})$, 
averaged over thermal fluctuations  
\begin{equation}
  \rho({\bf r}) = \av{\rho_\mu({\bf r})},
\end{equation}
where the brackets $\av{\dots}$ denote the thermal average. The 
liquid and solid phases are characterized by qualitatively 
different density fields $\rho({\bf r})$: in the liquid phase the particles are 
delocalized across the system and the averaged density 
$\rho({\bf r}) = \bar{\rho}^{\rss 3D}$ is constant;  on the other 
hand the solid phase is characterized by a modulated density 
$\rho({\bf r})$, with peaks at the lattice points.  
The appearance of finite density modulations is a consequence 
of particle-particle correlations arising from microscopic 
interactions. 

The classical DFT is based on the assumption 
that the free energy can be written as a functional of the 
averaged density $\rho({\bf r})$. One starts from the ideal gas 
free energy describing a non-interacting liquid and includes 
the correlations via an effective quadratic term in the density modulations 
$\delta \rho({\bf r}) =  \rho({\bf r})-\bar{\rho}^{\rss 3D}$. Within this 
approximation the grand canonical free energy difference relative to
the uniform liquid reads
\begin{align}
   \frac{\delta \Omega [\rho({\bf r})]} {T}\!
   = & \!\int\! d^3{\bf r} \, \Bigl[
   \rho({\bf r})\ln\frac{\rho({\bf r})}{\bar{\rho}^{\rss 3D}}
   -\delta \rho({\bf r}) 
   \nonumber \\
   & -\frac{1}{2}
   \int\!\!d^3{\bf r'}
   \delta \rho({\bf r})
   c(|{\bf r}\!-\!{\bf r}'|)
   \delta \rho({\bf r}')\Bigr], \label{func1a}
\end{align}
where the temperature $T$ is measured in unit of 
energy with $k_\mathrm{B} = 1$.
The first two terms generalize\cite{singh91} the
standard free energy of an ideal gas to the case of
non-homogeneous systems. The double integral term
incorporates the effects of interactions up
to second order in the density difference $\delta
\rho$.  This is the term which is responsible for the
appearance of finite density modulations which are 
absent in a non-interacting system. Therefore, the 
key input in this theory is the function $c(r)$, the so-called 
direct pair correlation function, which 
accounts for correlations in the reference liquid.

The direct pair correlation function can be related 
to more transparent physical quantities such as the 
static structure factor \cite{chaikinlubensky,hansenbook}. Here,
we give a brief derivation of the main relations which will be needed 
in our following discussion and, in particular, in the derivation 
of the Clausius-Clapeyron equation.
We start from the microscopic density-density correlator
\begin{align}\label{rhorho}
   \av{\delta \rho_\mu({\bf r}_1)\delta \rho_\mu({\bf r}_2)} \equiv
   \av{[\rho_\mu({\bf r}_1)\!-\!\bar{\rho}^{\rss 3D}]
   [\rho_\mu({\bf r}_2)\!-\! \bar{\rho}^{\rss 3D}]}.  
\end{align}
The Fourier transform of the density-density 
correlator defines the structure factor
\cite{note1}
\begin{displaymath}
   S({\bf q}) 
   = \frac{1}{\bar{\rho}^{\rss 3D}V}
      \int d^3{\bf r}_1\,d^3{\bf r}_2\, 
      e^{-i{\bf q}\cdot ({\bf r}_1-{\bf r}_2)}
       \av{\delta \rho_\mu({\bf r}_1)\delta \rho_\mu({\bf r}_2)}.
\end{displaymath}

Next, we calculate the structure factor from the free energy (\ref{func1a}).
The second functional derivative of the free energy $\delta \Omega$ 
with respect to $\rho({\bf r})$
evaluated at $\rho({\bf r}) = \bar\rho^{\rss 3D}$ is the (functional) inverse of 
the density-density correlator (see also Ref.\ \onlinecite{chaikinlubensky})
\begin{align}
  \frac{\delta^2 [\delta \Omega]}  
   {\delta [\delta \rho({\bf r}_1)] \delta [\delta \rho({\bf r}_2)]} & = 
   [ \av{\delta \rho_\mu({\bf r}_1)\delta \rho_\mu({\bf r}_2)}]^{-1} \nonumber \\
    & = \frac{1}{\bar\rho^{\rss 3D}} \delta({\bf r}_2 - {\bf r}_1) - 
    c(|{\bf r}_2 - {\bf r}_1|).
\end{align}
The functional inverse can be easily calculated in Fourier space. In this way, we 
find a relation between the structure factor and the direct 
correlation function\cite{note2}
\begin{equation}\label{sq}
   S({\bf q}) 
   = \frac{1}{1 - c({\bf q})}.
\end{equation}
Therefore, apart from additive constants (or delta functions in real space), 
the direct correlation function $c({\bf q})$ is given by the inverse of the 
structure factor $S({\bf q})$.
The $q=0$ mode of the structure factor is related through
the fluctuation-dissipation theorem to the isothermal
compressibility of the system, i.e., $S(q=0) 
=(\av{N^2} - \av{N}^2)/\av{N}^2 = \bar{\rho}^{\rss 3D}
T \kappa_T$ (the isothermal compressibility of the
ideal gas is $\kappa_T^0 = 1/(\bar{\rho}^{\rss 3D}T$))
and thus
\begin{equation}\label{ccomp}
  1-c(q=0) = (\bar{\rho}^{\rss 3D} T \kappa_T)^{-1}.
\end{equation}

The study of the free energy functional (\ref{func1a})
requires knowledge of the direct correlation function
$c(r)$,  a quantity which is usually obtained
from the liquid state theory. 
It is possible to write a diagrammatic expansion for $c(r)$
in terms of the microscopic two-body potential $V(r)$
(or more precisely of the Mayer function $f(r)=\exp(-V(r)/T)-1$). 
For weak potentials, 
high-order correlations can be neglected and $c(r)$ 
is given by the first (unperturbed) term\cite{hansenbook}
\begin{equation}\label{cflim}
  c(r) \approx f(r) \approx - V(r)/T.
\end{equation}
In order to obtain a better estimate for $c(r)$, higher order terms in the 
perturbation expansion must be included. This is usually 
done by selecting specific classes 
of diagrams out of the complete perturbative series. 
The approach that is most widely used
is the hypernetted chain (HNC) closure, an approximation
scheme which, however, is known to underestimate
liquid-state correlations.  The strategy we pursue
here is different: we exploit the specific properties 
of the pancake vortex lattice
by considering the system as a collection of two-dimensional 
systems of log-interacting particles
subject to a periodic modulated substrate potential
due to the other layers, as done in Ref.\ \onlinecite{dodgson00}. 
The function $c(r)$ then combines
an in-plane correlator, arising from the strong 
in-plane logarithmic interaction, and the weak out-of-plane potential.

\section{DFT-substrate approach}

In contrast to standard liquids, the pancake vortex
system exhibits a strong uniaxial anisotropy. Hence, 
we consider separately the in-plane (${\bf R}$) 
and out-of-plane ($z$) dependencies: in particular
$\rho({\bf r}) \rightarrow \rho_z({\bf R})$
becomes a sequence (in $z$) of two-dimensional densities.
Similarly, we define the density variations $\delta \rho_z({\bf R})
= \rho_z({\bf R}) - \bar{\rho}$ and the direct pair
correlation function $c_{z}(R)$
(note that $\bar{\rho}$ is a 2D density).  The DFT
free energy of Eq.\ (\ref{func1a}) can be adapted 
to the anisotropic vortex liquid
\begin{eqnarray}
   \lefteqn{\frac{\delta \Omega [\rho_z({\bf R})]} {T}\!
   = \int\! \frac{dz}{d} d^2{\bf R} \, \Bigl[
   \rho_{z}({\bf R})\ln\frac{\rho_{z}({\bf R})}{\bar{\rho}}
   -\delta \rho_{z} ({\bf R})} 
   &
   \nonumber\\
   && -\frac{1}{2}\!
   \int\!\! \frac{dz'}{d} d^2{\bf R'} 
   \delta \rho_{z}({\bf R})
   c_{z\!-\!z'}(|{\bf R}\!-\!{\bf R}'|)
   \delta \rho_{z'}({\bf R}')\Bigr]. \label{func1}
\end{eqnarray}
The only input needed in the DFT free energy
is the direct correlation function $c_{z}(R)$ which 
we obtain by implementing the substrate model
for the pair correlation function,
\begin{equation}\label{csub}
   c_{z}(R) =  d c^{\rss 2D}(R) \delta(z) - \frac{V_{z}(R)}{T},
\end{equation}
where $V_{z}(R)$ is the out-of-plane interaction of Eq.\ (\ref{vz}). 

Within the planes, vortices are strongly correlated due to the 
repulsive logarithmic interactions (Eq.\ (\ref{v0})). 
Hence, we can approximate 
$c_{0}(R)$ with the direct correlation function $c^{\rss 2D}(R)$ 
of the two dimensional logarithmically interacting particles 
(also known as one-component plasma, OCP).
The in-plane component also contains contributions 
from the out-of-plane interactions $V_{z\neq 0}(R)$, which 
are small however, since $V_{z\neq 0}(R)$ appears in the perturbative expansion
\cite{hansenbook} of $c_0(R)$ at least to quadratic order, 
$[V_{z\neq 0}(R)/T]^2 \sim (d/\lambda)^2$. Hence, the overall
contribution due to out-of-plane interactions of all planes adds
up to $\sim (\lambda/d)[V_{z\neq 0}(R)/T]^2 \sim (d/\lambda)$. 

We use results of Monte Carlo simulations of the
two-dimensional  OCP at various coupling constants
$\Gamma = 2\varepsilon_0 d/T$ to extract $c^{\rss
2D}(R)$. We have performed simulations on a
system of 256 particles, using an alternative to
the traditional Ewald summation method proposed
recently by Tyagi\cite{Tyagi2004}. Thermodynamic
data and correlations were averaged over $\sim 3
\times 10^3$ independent measurements following
equilibration. The program was benchmarked using
available numerical results for correlations and
thermodynamic functions. In the simulations, the structure factor 
$S(K)$ was calculated to yield the direct correlation 
function via the relation $c(K)= 1-1/S(K)$, see Eq.\ (\ref{sq}).

In the determination of the out-of-plane direct
correlation function $c_{z\neq 0}(R)$ we neglect the
higher order terms in the potential expansion, approximating it with
the leading unperturbed value $-V_{z \neq 0}(R)/T$
(cf. Eq.\ (\ref{cflim})).  Higher orders in $c_{z\neq 0}(R)$ 
involve at least terms of order $\sim(d/\lambda)^2$. 
In the following, we will see that the relevant 
quantity in our analysis is the total out-of-plane correlator, 
defined as $\int (dz/d)c_{z\neq 0}(R)$. Whereas the leading 
term is of order $(\lambda/d)(d/\lambda)\sim 1$ and hence 
comparable to the in-plane component, the subleading term
is of order $(\lambda/d)(d/\lambda)^2\sim (d/\lambda)$. 
We neglect this contribution to be consistent with our approximation 
for the in-plane component of the correlator.

At a mean-field level the thermodynamically stable state
corresponds to the minimal free energy configuration of the
functional (\ref{func1}). Then, the density functions
$\rho_z(R)$ must obey the saddle point
equation
\begin{equation}\label{min1}
  \ln \frac{\rho_z({\bf R})}{\bar{\rho}} = \int
  \frac{dz'}{d}\int d^2{\bf R'} \, c_{z-z'} (|{\bf
  R}\! -\! {\bf R'}|) \delta \rho_{z'}({\bf R'}).
\end{equation} 
A key quantity in our discussion is the molecular 
field\cite{ramakrishnan79,ramakrishnan82,chakrabarti94}
 $\xi_z({\bf R})$ defined through
\begin{equation}\label{molfield}
  \xi_z({\bf R}) = \ln (\rho_z({\bf R})/\bar{\rho}).
\end{equation}
At the minimum of the free energy, combining the saddle point equation 
(\ref{min1}) with (\ref{molfield}), the molecular field becomes
\begin{equation}\label{molfield2}
  \xi_z({\bf R}) =  \int \frac{dz'}{d}\int d^2{\bf R'} \, c_{z-z'}
  (|{\bf R}\! -\! {\bf R'}|)
  \delta \rho_{z'}({\bf R'}).
\end{equation}
Hence, the molecular field represents the screening potential, produced
by the modulated density which is acting back on the density itself.  
However, while (\ref{molfield}) defines the molecular
field and hence applies always, the interpretation in terms of a screening
potential following from Eq.\ (\ref{molfield2}) is valid only 
at the minimum.

\section{Free energy in Fourier space} 

\begin{figure}[t]
\centering
   \includegraphics [width= 7.5cm] {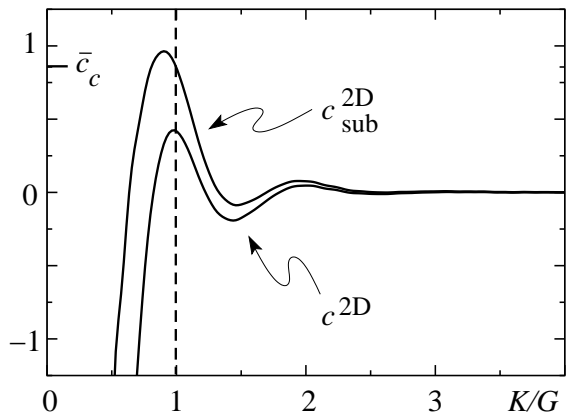}
   \caption
   [Direct correlation functions: OCP plasma and OCP with
   substrate]
   {Direct correlation function at $T/ \varepsilon_0 d =0.1$
   ($\Gamma = 20$) for the two-dimensional OCP, 
    $c^{\rss 2D}(K)$ from MC simulations and  
   $c^{\rss 2D}_{\rss sub}(K)$ (cf.\ Eq.\ (\ref{c3d_s}))
   for the full three-dimensional pancake vortex system 
   at the melting field $B_\mathrm{m}$ 
   ($B_\mathrm{m}/B_\lambda \approx 0.099$ for 
   $T/ \varepsilon_0 d =0.1$). 
   We define our wave-length unit as 
   $G \equiv (8\pi^2\bar\rho/\sqrt{3})^{1/2}$. 
   The substrate potential $\tilde{V}_{\rss stack}(K)$ modifies 
   the full correlator $c^{\rss 2D}_{\rss sub}(K)$ in two different 
   ways as compared to the in-plane correlation function 
   $c^{\rss 2D}(K)$: {\it i)} it enhances the correlations 
   of the liquid, pushing the melting to temperatures larger than 
   $T^{\rss 2D}_\mathrm{m}$ and {\it ii)} it shifts the peak of the 
   correlation function to a value of $K$ which is smaller than 
   $G$. In the incompressible limit $K_1 = G$ and at melting
   the full three-dimensional correlator assumes the critical value
   $\bar{c}^{\rss 2D}_{\rss sub} \equiv {c}^{\rss 2D}_{\rss sub}(G) 
   = \bar{c}_c \approx 0.856$, cf.\ (\ref{critical_inc}).
   Considering a finite compressibility, the system is allowed to gain 
   correlation energy by
   crystalizing at  $K_1 < G$, hence, at a density smaller 
   than that of the liquid.
   However, a large density change is prevented by the finite 
   compressibility of the system and the crystalized structure is 
   characterized by a first reciprocal lattice vector
   $K_1\lesssim G$ (see text).} \label{fig:c3d}
\end{figure}

In thermodynamic equilibrium all superconducting planes
are equivalent and the averaged vortex density 
$\rho_z({\bf R})$ becomes independent of the layer
position $z$; we write $\rho_z({\bf R}) = \rho({\bf R})$.
Next, instead of seeking the exact form
$\rho({\bf R})$ solving the non-linear integral
equations (\ref{min1}), we restrict our analysis to
a simple family of periodic functions which model
the modulations of the density in the triangular crystalized phase.
In the following, we concentrate on the simplest case,
retaining only the first Fourier components of the
density in a triangular lattice
\begin{equation}\label{ansatzdft}
    \frac{\rho({\bf R})}{\bar{\rho}} = 1 + \eta +
    \sum_{{\bf K}_1} \mu e^{i {\bf K}_1\cdot {\bf R}}
    = 1 + \eta +\mu g_{K_1}({\bf R}),
\end{equation} 
where the vectors ${\bf K}_1$
are the first reciprocal lattice vectors of the
frozen structure and depend on the area $a$ of the 
unit cell, $\mu = \delta \rho(K_1)
/\bar{\rho}$ is the Fourier component of the
density with wave length $K_1$, and $\eta =
\delta \rho({\bf K} = 0)/\bar{\rho}$ is the relative
density change upon freezing. 
Within this approach the density is characterized 
by the three variables $\mu$, $\eta$, and $K_1$. Note that, however, 
in a scheme where the number of particles in a unit cell is constrained
to be an integer, the size of the unit cell $a$ in the 
crystalized structure and the value of the 
density jump are related and, thus, 
$K_1$ and $\eta$ are not independent variables; we return to
this issue in the next section. The function
\begin{equation}\label{funcg}
   g_{K_1}({\bf R}) = \sum_{{\bf K}_1} 
   e^{i {\bf K}_1 \cdot {\bf R}}= 2 \cos (2\tilde{x}) 
   + 4 \cos(\tilde{x})\cos(\tilde{y})
\end{equation}
includes the sum over the six first reciprocal vectors in the triangular 
lattice; in the last equality we have defined 
the dimensionless variables $\tilde{x} = x K_1/2$ and 
$\tilde{y} = \sqrt{3}\, yK_1/2$.
We also write a similar Ansatz for the molecular field
\begin{equation}\label{ansatzxi}
  \xi({\bf R}) = \zeta +  \xi g_{K_1}({\bf R}),
\end{equation}
retaining only the zeroth and first Fourier
components, $\zeta$ and $\xi$ respectively, consistent
with (\ref{molfield2}) and the rapid decay of the correlator
$c(K)$ in the liquid phase, cf.\ Fig.\ \ref{fig:c3d}.

The Fourier components of $\rho({\bf R})$ and $\xi({\bf R})$
are not independent and can be related through (\ref{molfield}). 
With the help of the relations 
\begin{align}
  \frac{1}{a}\int_a d^2{\bf R}g({\bf R}) = 0, \quad\quad
   \frac{1}{a}\int_a d^2{\bf R}[g({\bf R})]^2 = 6,
\end{align}
we project the zeroth and first Fourier components of 
$\rho({\bf R}) = \bar \rho \exp (\zeta + \xi g_{K_1}({\bf R}))$ 
and obtain
\begin{align}
  \zeta & = -\Phi(\xi) + \ln (1+\eta), \nonumber \\
  \mu &   = \frac{1+\eta}{6}\, \Phi'(\xi),\label{ximu}
\end{align}
where we have defined the function
\begin{equation}\label{phixi}
  \Phi(\xi) = \ln \left[ 
  \frac{1}{a}\int_a d^2{\bf R}\, e^{\xi g({\bf R})}
  \right].
\end{equation}
This function is unaffected by rescaling  
of the unit cell area $a$ (equivalently, it
does not depend 
on $K_1$).

Substituting the Ans\"atze (\ref{ansatzdft}) and (\ref{ansatzxi})
in the DFT free energy (\ref{func1}), we obtain the 
two-dimensional free energy
density
  $\delta \omega^{\rss 2D}_{\rss sub}\equiv
  (d/\bar{\rho} V)\,\delta \Omega$
as a function of the order parameters $\eta$ and $\mu$ and the 
length $K_1$ of the first reciprocal lattice vectors,
\begin{align}\label{freeen_ansatz}
  \frac{\delta \omega^{\rss 2D}_{\rss sub}(\eta,\mu,K_1)}{T} & =  
   (1+\eta)[\ln (1+ \eta) - \Phi(\xi)] -\eta \nonumber \\
  & + 6\xi \mu  -\frac{c^{\rss 2D}_{\rss sub}(0)\eta^2}{2} 
  - 3 c^{\rss 2D}_{\rss sub}(K_1)\mu^2, 
\end{align}
where $\xi$ has to be understood as a function of $\eta$ 
and $\mu$ through (\ref{ximu}). In (\ref{freeen_ansatz}) we 
have defined the correlator 
\begin{equation}\label{c3d_s}
   c^{\rss 2D}_{\rss sub}(K) =  \int \frac{dz}{d}\, c_z (K) =
   c^{\rss 2D}(K) -  \int \frac{dz}{d}\, \frac{\tilde V_z (K)}{T} ,
\end{equation}
where $\tilde V_z(K) = \bar\rho V_z(K)$ 
is the dimensionless Fourier transform (with an additional $\bar\rho$ factor) 
of the out-of-plane pancake vortex interaction (Eq.\ (\ref{vk})). 
The out-of-plane interactions contribute to the correlator 
through the total stack potential 
\begin{align}\label{vstackg}
  - \frac{\tilde{V}_{\rss stack}(K)}{T}\! = \!-\!\!
   \int_{-\infty}^{+\infty} \!
   \frac{dz}{d} 
   \frac{\tilde V_{z}(K)}{T} = \!\frac{4 \pi  \bar\rho \varepsilon_0 d}
  {TK^2 (\lambda^2K^2 \!+ 1)},
\end{align}
which enhances the nearest-neighbor correlations, cf. Fig.\ \ref{fig:c3d}.

Within this approach only the $K=0$ and the $K = K_1$ components 
of the correlator are present in the expression for the free energy 
(\ref{freeen_ansatz}).  The $K=K_1$ component measures the
correlations which are responsible for the solidification of the 
liquid; its dependence on temperature and field yields the
position of the melting line and its local dependence on $K$ 
determines the discontinuities in the first order transition.
On the other hand, the $K=0$ component $c^{\rss 2D}_{\rss sub}(0)$ 
of the correlator is related to the compressibility of the vortex 
system via (\ref{ccomp}). Using the relation $\kappa_T = 1/c_{11}(0)$ 
between the compressibility and the compression modulus and using the
expression\cite{brandt77} $c_{11}(0) = B^2/4\pi$, we find that 
\begin{equation}\label{c3d0}
  1- c^{\rss 2D}_{\rss sub}(0) = \frac{ 4\pi \varepsilon_0 d}{T}\, 
  \frac{B}{B_\lambda} =
  \frac{\Phi_0 B d}{4\pi T},
\end{equation}
consistent with the result of the present substrate-based approach: 
here, the $K \to 0$ divergence in the correlator $c^{\rss 2D}(K\to 0) 
\sim -4 \pi \bar\rho \varepsilon_0d/TK^2$ of the incompressible Coulomb 
gas\cite{caillol82} is cancelled by the corresponding divergence in 
the out-of-plane component of the correlator, see (\ref{vstackg}), 
and the remaining term reproduces (\ref{c3d0}). In the Coulomb gas 
problem, this cancellation has to be achieved by the introduction 
of a compensating background.

Our expression for the free energy (\ref{freeen_ansatz}) in terms of 
the density  $\rho$ has to be compared with the original formula in Ref.\ 
\onlinecite{ramakrishnan79}, where the free energy has been given 
in terms of the molecular field $\xi$. The two approaches are related 
via a Legendre transformation which adds the energy of an external 
periodic potential $u(r)$, $\delta W(u) = \min_{\rho}[\delta \Omega(\rho) 
- \int d^3 r\, u(r) \rho(r)]$. The relation (\ref{molfield2}) then 
includes the external potential $u$, $\xi_z(R) = u_z(R) + (1/d)\int dz'
\int d^2 {\bf R'} c_{z-z'}(|{\bf R}-{\bf R'}|)\delta \rho_{z'}(R')$. 
Setting $u=0$, both approaches provide identical results.
However, the formulation given here is more appropriate when describing
non-uniform configurations as they occur in a solid-liquid interface
or near a surface, cf.\ Ref.\ \onlinecite{decol05b}. 

\section{Constrained Theory}\label{sec:fa}

Within our approximation for $\rho(R)$, the state of the 
system is characterized by three parameters:  
the density modulation $\mu$ at the first reciprocal lattice vector, 
the density change $\eta$ across the transition, and the wave number 
$K_1$. However, a theory based on the Ansatz (\ref{ansatzdft})
and the functional (\ref{func1}), in which $\eta$, $\mu$, and 
$K_1$ are independent variables, is not fully consistent. The problem,
as was pointed out in Ref.\ \onlinecite{laird87}, is that in 
(\ref{ansatzdft}) the density jump $\eta$ and the wave number 
$K_1$ appear as independent variables. This theory therefore 
may lead to the appearance of states with a non-integer 
occupancy per unit cell and hence to an incorrect
description of the solidification of the liquid\cite{note_con}. To solve this inconsistency 
only states with a fixed integer total number of particles per unit cell
should be considered. One must then proceed with a constrained
minimization of the free energy, introducing a Lagrange
multiplier $\chi$ which enforces the so-called `perfect crystal' condition 
\cite{laird87}
\begin{equation}\label{constraint1}
	\int_a d^2{\bf R}\, \rho({\bf R}) = 1,
\end{equation}
i.e., each unit cell contains exactly one vortex.
With this additional term the free energy density becomes
\begin{align}
   \frac{\delta\tilde{\omega}^{\rss 2D}_{\rss sub} (\eta,\mu,K_1,\chi)}{T}
   = & \frac{\delta \omega^{\rss 2D}_{\rss sub}(\eta,\mu,K_1)} {T} 
   \nonumber \\
   & \quad - 
   \frac{\chi}{\bar{\rho}a} 
   \Bigl[ \int_a\! 
   d^2{\bf R}\, \rho({\bf R}) -1\Bigr],
   \label{funcl}
\end{align}
where $\delta \omega^{\rss 2D}_{\rss sub}(\eta,\mu,K_1)$ is given by
(\ref{freeen_ansatz}). Substituting the Fourier Ansatz in the Lagrange 
multiplier term, we obtain the expression for the constrained 
free energy
\begin{align} 
  \frac{\delta \tilde{\omega}^{\rss 2D}_{\rss sub}(\eta,\mu,K_1,\chi)}{T} 
  & =  \frac{\delta \omega^{\rss 2D}_{\rss sub}(\eta,\mu,K_1)}{T}\nonumber \\ 
  & \quad - 
   \chi \Bigl[ (1+\eta)- \Bigl(\frac{K_1}{G}\Bigr)^2\Bigr],
   \label{freeen_constr}
\end{align}
where $G= (8\pi^2\bar{\rho}/\sqrt{3})^{1/2}$ is the length of the first 
reciprocal lattice vector associated with a solid with the same density 
as the liquid. 

To obtain the constrained minimum of the free energy, one needs to 
solve the saddle point equations for the variables $\chi$, $\mu$, $\eta$, and $K_1$.
Taking the derivative with respect to the Lagrange multiplier $\chi$, 
we recover the `perfect crystal' constraint (\ref{constraint1}) in the form 
\begin{equation}\label{etak1}
  \eta(K_1) = \left( \frac{K_1}{G} \right)^2 -1,
\end{equation}
which yields a relation between $\eta$ and $K_1$. 
The case $K_1=G$ describes an incompressible system, 
since the solid and the liquid have the same density.
However, an ordinary first order phase transition is characterized 
by a finite jump of the density, and hence by a non-zero $\eta$.
Consistent with our constrained theory, a finite $\eta$ 
corresponds to the crystallization into a solid with a vortex density
$n_\mathrm{v}\neq \bar\rho$ and, hence, with a first reciprocal 
lattice vector $K_1 \equiv (8\pi^2 n_\mathrm{v}/\sqrt{3})^{1/2}$ 
which is different from $G$, $K_1\neq G$. When $K_1 > G$ the solid 
is denser than the liquid, as in conventional materials, 
whereas $K_1 < G$ leads to an anomalous density jump with a solid 
which is less dense than the liquid, as is the case in the 
water-ice transition. 

Next, the extremum condition of the free energy 
(\ref{freeen_constr}) with respect to $\mu$ gives the relation
\begin{equation}
   \xi = c^{\rss 2D}_{\rss sub}(K_1) \mu, \label{min_constr1}
\end{equation}
which is essentially Eq.\ (\ref{molfield2}) in Fourier space. The equation 
for $\eta$
\begin{equation}   
   \chi = -\Phi(\xi) + (1-c^{\rss 2D}_{\rss sub}(0))\eta,\label{min_constr2}
\end{equation}
is modified as compared with the standard unconstrained theory by the 
presence of the Lagrange multiplier, which can be found from the
minimization of (\ref{freeen_constr}) as a function of $K_1$
\begin{equation}  
   \chi = \frac{3 G^2 \mu^2}{2 K_1} \frac{\partial 
   c^{\rss 2D}_{\rss sub}(K)}{\partial K}\Bigr|_{K_1}.
   \label{min_constr3}
\end{equation}
Combining (\ref{min_constr1})-(\ref{min_constr3}) and the relation
between $\mu$ and $\xi$ as given by (\ref{ximu}),
we can eliminate $\xi$ and $\chi$. The saddle point equations
for $\eta$ and $\mu$ can be written as
\begin{align}\label{min2c}
  \mu  & = 
  \frac{[1+\eta]}{6} \frac{\ds \int_a d^2{\bf R}\, g({\bf R}) 
  \exp\Bigl[ \mu c^{\rss 2D}_{\rss sub}(K_1) g({\bf R})\Bigr]}
 {\ds \int_a d^2{\bf R} 
 \exp\Bigl[\mu c^{\rss 2D}_{\rss sub}(K_1) g({\bf R})\Bigr]}, \\
  \noalign{\vskip 5 pt}
 \eta & = \frac{1}{1-c^{\rss 2D}_{\rss sub}(0)}
 \left[ \Phi(c^{\rss 2D}_{\rss sub}(K_1)\mu) + \right. \nonumber \\
 & \qquad\qquad\qquad
 \left.\frac{3 G^2 \mu^2}{2 K_1} \frac{\partial c^{\rss 2D}_{\rss sub}(K)}{\partial K}
 \Bigr|_{K_1}
  \right],\label{min2c_2}
\end{align}
where $\eta$ and $K_1$ are related by (\ref{etak1}).  
The homogeneous ($\mu= 0$) and uncompressed ($\eta = 0$) 
liquid always solves these equations.
However, at low temperatures, other non-uniform
solutions ($\mu \neq 0$) may appear. If their corresponding
free energy is smaller than that of the liquid, $\delta \omega^{\rss 2D}_{\rss sub} < 0$,
the system freezes into the periodic (crystal) structure (note that at the 
minimum the perfect crystal constrained is fulfilled and 
$\delta \tilde{\omega}^{\rss 2D}_{\rss sub} = \delta \omega^{\rss 2D}_{\rss sub}$).

The equations (\ref{min2c}) and (\ref{min2c_2}) can also be obtained directly 
from the free energy (\ref{freeen_ansatz}),
\begin{equation}
 \delta \omega^{\rss 2D}_{\rss sub}(\eta, \mu)\equiv
 \delta \omega^{\rss 2D}_{\rss sub}(\eta,\mu,K_1(\eta)),
\end{equation}
where $K_1$ is written as a function of $\eta$ via (\ref{etak1}).
The minimization of  $\delta \omega^{\rss 2D}_{\rss sub}(\eta, \mu)/T$
with respect to $\eta$ and $\mu$ then yields the saddle point equations 
(\ref{min2c}) and (\ref{min2c_2}).

\section{Incompressible limit and Melting line}\label{sec:melting_in}

The pancake vortex system is essentially incompressible 
($|c^{\rss 2D}_{\rss sub}(0)|\gg 1$)
in a wide portion of the phase diagram and the corresponding 
value of the density jump is small, i.e., $\eta\ll 1$. 
Hence, the determination of the melting line can be done within
the incompressible limit, with $\eta = 0$ and $K_1 = G$. This 
approximation is sufficiently accurate also for small magnetic fields 
(see later for estimates of the value of $\eta$). In the incompressible limit, the free energy becomes 
a function of $\mu$ alone,
\begin{align}
  \frac{\delta \omega^{\rss 2D}_{\rss sub}(\mu)}{T} & 
  \equiv\frac{\delta \omega^{\rss 2D}_{\rss sub}(\eta=0,\mu)}{T} \nonumber \\
  &  =
     6\xi(\mu) \mu - 3\bar{c}^{\rss 2D}_{\rss sub}  \mu^2 
     - \Phi(\xi(\mu)),\label{free2d0}
\end{align}
where $\bar{c}^{\rss 2D}_{\rss sub}\equiv c^{\rss 2D}_{\rss sub}(G)$ and $\xi(\mu)$
is implicitly defined by
\begin{equation}\label{muxi}
  \mu(\xi) = \Phi'(\xi)/6.
\end{equation}
\begin{figure}[t]
  \centering
   \includegraphics [width= 7.8cm] {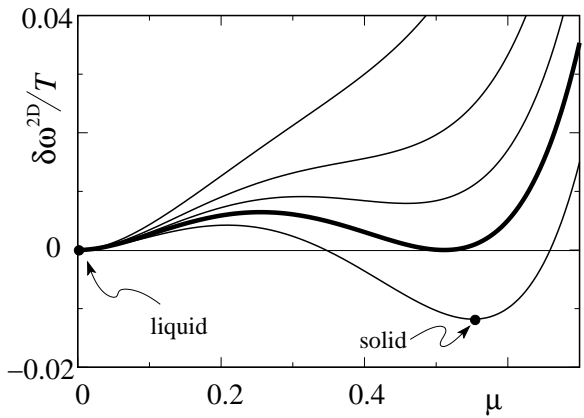}
   \caption
   [Two dimensional free energy density as a function of the 
   order parameter $\mu$]
   {Profiles of the two dimensional free energy difference 
   $\delta \omega^{\rss 2D}(\mu)$ of Eq.\ (\ref{free2d0}) 
   as a function of the order parameter $\mu$, 
   in correspondence to the values (from above to below) 
   $\bar{c}^{\rss 2D} \equiv c^{\rss 2D} (G)$ 
   = 0.80, 0.83, 0.845, 0.856 ($=\bar{c}_c$ critical, 
   thicker line), 0.87. At melting the order parameter jumps
   from the solid minimum at $\mu_\mathrm{s} \approx 0.51$
   to $\mu_\mathrm{l} = 0$, overcoming the barrier
   $\delta \omega_{\rss max}^{\rss 2D} \approx 0.0065\, T$.}
\label{fig:f2d}
\end{figure}

The effective three-dimensional correlator $\bar{c}^{\rss 2D}_{\rss sub}$ entering (\ref{free2d0}) is given by the sum 
of two contributions: the OCP correlation function $\bar{c}^{\rss 2D}
\equiv c^{\rss 2D}(G)$ and the stack potential 
$\tilde{V}_{\rss stack}(G)/T$ of Eq.\ (\ref{vstackg}), both 
evaluated at $K = G$. 
While the first depends on temperature only, the latter 
depends also on the vortex density and thus on the magnetic field,
\begin{align}\label{c2dbt}
  \bar{c}^{\rss 2D}_{\rss sub}(T,B) & =  \bar{c}^{\rss 2D}(T) 
     + \frac{4\pi\bar{\rho}\varepsilon_0d}{TG^2}\frac{1}{1+\lambda^2 G^2}
     \\
  & = \bar{c}^{\rss 2D}(T) + \frac{\sqrt{3}\varepsilon_0d}{2 \pi T} 
  \frac{1}{[1+ (8\pi^2/\sqrt{3})B/B_\lambda] } \nonumber, 
\end{align}
where we use $\bar{\rho}/G^2 = \sqrt{3}/(8\pi^2)$ and 
$\lambda^2 G^2= (8\pi^2/\sqrt{3})$
$B/B_\lambda$.

\begin{figure}[t]
\centering
   \includegraphics [width= 7.8 cm] {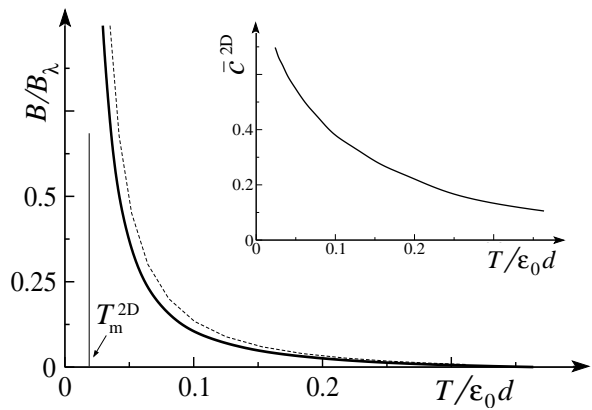}
   \caption
   [DFT Melting line: comparison with full numerical solution]
   {Comparison of the melting line obtained from the 
   substrate-DFT analysis  (solid line) with the result of full 
   numerical simulations \cite{dodgson00} (dashed line).
   Inset: values of the first peak at $K_\mathrm{max}\approx G$ of the 
   in-plane two-point direct correlator $c^{\rss 2D}(K)$ as a function of 
   $T/\varepsilon_0 d = 2 / \Gamma$, from Monte Carlo 
   simulations of the 2D one component plasma (OCP).
}   
   \label{fig:dftbulk}
\end{figure}

At large values of $B$, the inter-plane interaction is negligible
and the full correlator reduces to the 2D-OCP component 
$\bar{c}^{\rss 2D}$.  
The temperature enters via the $T$-dependence of the 
direct correlation function $\bar{c}^{\rss 2D}(T)$, changing
the coefficient of the quadratic term (as in the $\phi^4$ 
Landau theory). We obtain $\bar{c}^{\rss 2D}(T)$ directly from MC 
simulations; the results are shown in the inset of Fig.\ \ref{fig:dftbulk}. 
Increasing the temperature, the liquid correlations weaken,
$S(G)$ decreases, and so does $\bar{c}^{\rss 2D}$, 
cf.\ Eq.\ (\ref{sq}). 

As a function of $\mu$, the free 
energy exhibits the shape of a Landau theory describing 
a first-order phase transition. In Fig.\ \ref{fig:f2d} we plot
the free energy as a function of $\mu$ for different
values of $\bar{c}^{\rss 2D}$. 
At large temperature, the correlator $\bar{c}^{\rss 2D}$ is small and 
$\delta \omega^{\rss 2D}(\mu)$ exhibits only one minimum
at $\mu = 0$ with a value $\delta \omega^{\rss 2D}(0)/T = 0$, 
in correspondence with the (homogeneous) liquid
phase. Decreasing the temperature (which corresponds to
increasing $\bar{c}^{\rss 2D}$), a second local 
minimum $\mu_\mathrm{s}$ (metastable solid) with energy
\begin{equation}
   \frac{\delta \omega^{\rss 2D}(\mu_\mathrm{s})}{T} 
    = 3\bar{c}^{\rss 2D}  \mu_\mathrm{s}^2 
     - \Phi(\bar{c}^{\rss 2D}\mu_\mathrm{s})\label{free2d0_min}
\end{equation}
appears in addition to the liquid minimum at $\mu_\mathrm{l}=0$.
Freezing occurs when the liquid and solid minima assume 
the same value of the free energy, i.e., when 
$\delta \omega^{\rss 2D}(\mu_\mathrm{s}) = 0$.
Within our single order parameter theory, this 
condition is equivalent to a simple equation for the 
correlator \cite{ramakrishnan82}
\begin{equation}\label{critical_inc}
  \bar{c}^{\rss 2D} = \bar{c}_c \approx 0.856.
\end{equation}
Going to even lower temperatures, $\bar{c}^{\rss 2D}$ further 
increases, the solid minimum decreases in value, 
$\delta \omega^{\rss 2D}(\mu_\mathrm{s})/T < 0$,
and the crystal becomes the only thermodynamically stable phase.
Monte Carlo simulations \cite{caillol82} show that the 2D-OCP
freezes at $T_\mathrm{m}^{\rss 2D} \approx \varepsilon_0d/70$
where, however, the correlator assumes the value $\bar{c}^{\rss 2D} 
\approx 0.77 <\bar{c}_c$ ($\Gamma^{\rss 2D}_\mathrm{m} = 
2\varepsilon_0d/T^{\rss 2D}_\mathrm{m}=140$). 
This disagreement 
is due to the approximations we have adopted in 
our analysis. In particular, at low temperatures, the higher-order 
peaks become important and more terms in the Fourier expansion 
have to be retained \cite{singh91}. 

At lower magnetic fields, the inter-plane correlation becomes 
important and the 2D correlations $\bar{c}^{\rss 2D}$ are 
augmented by the stack potential $V_{\rss stack}(G)$. 
The critical condition $\bar{c}^{\rss 2D}_{\rss sub}(T,B)=\bar{c}_c$ 
can be solved together with (\ref{c2dbt}) and yields
a simple expression for the melting line $B_\mathrm{m}(T)$
\begin{equation}\label{bt_bulk}
  \frac{B_\mathrm{m}(T)}{B_\lambda} = 
  \frac{\sqrt{3}}{8\pi^2}
    \Bigl[
  \frac{\sqrt{3}\Gamma}
  {4\pi (\bar{c}_c-\bar{c}^{\rss 2D}(T))} -1 
  \Bigr].
  \end{equation}
This melting line is plotted in Fig.\ \ref{fig:dftbulk} (lower) together
with the numerical results of the MC/MD simulations\cite{fangohr03}.
Furthermore, we find improved agreement in comparison with previous 
DFT studies where the direct correlation function
was derived {\it ab initio} through approximative closure schemes 
such as the hypernetted chain or the more 
elaborate Rogers-Young approach\cite{sengupta91,menon96}. In 
particular our novel approach approximates well the numerical 
results for small values of the magnetic fields well
($B\lesssim 0.5\,B_\lambda$), in a regime where 
results from previous work exhibit a substantial disagreement.

%

\section{Density jump and Clausius-Clapeyron relation}

To quantify the density jump across
the transition, we need to consider the saddle 
point equation (\ref{min2c_2}) for $\eta$
\begin{align}
  \eta & = \frac{1}{1-c^{\rss 2D}_{\rss sub}(0)}
  \Bigl[ \Phi(c^{\rss 2D}_{\rss sub}(K_1) \mu) \nonumber \\\
 & \qquad\qquad\qquad  + \frac{3 G\mu^2}
  {2(1+\eta/2)}
  \frac{\partial c^{\rss 2D}_{\rss sub}(K)}{\partial K}\Bigr|_{K_1}
 \Bigr],
     \label{etacomp}
\end{align}
where we have used $K_1 = G\sqrt{1+\eta} \approx G(1+\eta/2)$ 
to linear order in $\eta$ from (\ref{etak1}).
Previous analyses\cite{cornaglia00} of the freezing of the pancake vortex
system were based on the unconstrained free energy (\ref{freeen_ansatz}).
In the unconstrained theory the first reciprocal lattice vector is fixed at $K_1 = G$ 
and the equation for $\eta$ contains only 
the first term in (\ref{etacomp}), since the second term is due to  
the Lagrange multiplier, cf.\  (\ref{min_constr3}). Hence, within the 
unconstrained theory, one obtains (`nc' stands for not-constrained)
\begin{equation}\label{etanc}
 \eta_\mathrm{nc} = \frac{\Phi(\bar{c}^{\rss 2D}_{\rss sub} \mu_\mathrm{s})}
{1- c^{\rss 2D}_{\rss sub}(0)}= \frac{4\pi T 
\Phi(\bar{c}^{\rss 2D}_{\rss sub} \mu_\mathrm{s})}{\Phi_0 B d}, 
\end{equation}
where we used (\ref{c3d0}) for $ c^{\rss 2D}_{\rss sub}(0)$. 
Given that 
$\Phi(\bar{c}^{\rss 2D}_{\rss sub} \mu_\mathrm{s}) >0$, the sign of 
$\eta$ is always positive. This result contradicts the 
Clausius-Clapeyron relation and experimental 
evidence that vortices, like water, freeze into a solid which is 
less dense than the liquid. In a magnetic system the 
Clausius-Clapeyron relation reads \cite{vortexchapter} 
\begin{equation}\label{cc}
  \Delta B = -4\pi \Delta s 
  \Bigl( \frac{dH_\mathrm{m}(T)}{dT}\Bigr)^{-1},
\end{equation}
where $\Delta B = B_\mathrm{l} - B_\mathrm{s} 
= -\eta B_\mathrm{l} = -\eta \Phi_0 \bar{\rho}$ is the jump in magnetic 
induction and $\Delta s = \Delta S/V = (S_\mathrm{l} - S_\mathrm{s})/V$ 
the (positive) 
jump in entropy density on heating. Ignoring the small difference between
$H$ and $B$, equation (\ref{cc}) can be rewritten as (`CC' stands 
for Clausius-Clapeyron)
\begin{equation}\label{cc1a}
  \eta_\mathrm{CC} = \frac{4\pi \Delta s}{\Phi_0 \bar\rho} 
  \Bigl( \frac{dB_\mathrm{m}(T)}{dT}\Bigr)^{-1}.
\end{equation}
Combining (\ref{cc1a}) with the negative slope 
of the melting line $B_\mathrm{m}(T)$ of Eq.\ (\ref{bt_bulk}), we obtain that 
the density jump on heating is positive, which corresponds to a 
negative $\eta$, thus $\eta_\mathrm{CC} < 0$. Therefore, a theory 
with a positive $\eta$  and a melting line with a negative slope is 
not thermodynamically consistent.  

The second term in (\ref{etacomp}) resolves this inconsistency.
In 2D systems the first maximum of the direct correlation
function shows up at $K_\mathrm{max} \approx G$, and hence 
$\partial_K c^{\rss 2D}_{\rss sub}(K)|_G$ (and the second term in (\ref{etacomp})) 
is zero and $\eta > 0$. However, for the 3D pancake vortex system the substrate potential 
shifts the first peak of $c^{\rss 2D}_{\rss sub}(K)$ 
to a value of $K$ which is smaller than $G$, 
$K_\mathrm{max} < G$ (cf.\ Fig.\ \ref{fig:c3d}). The system gains
correlation energy by crystalizing at a $K_1 < G$, hence, at a
density which is smaller than that of the liquid. However, a large 
density change is prevented by the finite compressibility of the 
system and the crystalized structure is characterized by a first reciprocal 
lattice vector $K_1$ below but still close to $G$, $K_1\lesssim G$.
Hence, in (\ref{etacomp}), the derivative 
$\partial_K c^{\rss 2D}_{\rss sub}(K)|_{K_1}$ at $K_1$ is negative, 
$\partial_K c^{\rss 2D}_{\rss sub}(K)|_{K_1}
\approx  \partial_K c^{\rss 2D}_{\rss sub}(K)|_G <0$, see Fig.\ \ref{fig:c3d}. 
The second term of  (\ref{etacomp}) is therefore negative and a negative 
solution for $\eta$ becomes possible.  

Next, we confirm the thermodynamic 
consistency of the constrained theory by showing that (\ref{etacomp}) 
and the Clausius-Clapeyron equation (\ref{cc1a}) are equivalent. 
To compare (\ref{cc1a}) with (\ref{etacomp}), we need to calculate the jump 
$\Delta s$ in entropy density in (\ref{cc1a}). The latter is given by 
the temperature derivative of the free energy difference 
\begin{align} 
  \Delta s & = \frac{\bar{\rho}}{d} 
  \frac{\partial \delta \omega^{\rss 2D}_{\rss sub}}{\partial T} 
  \Bigr|_{\eta,\mu} = \frac{\bar{\rho}T}{d} 
  \frac{\partial}{\partial T}\frac{\delta \omega^{\rss 2D}_{\rss sub}}{T} 
  \Bigr|_{\eta,\mu} \nonumber \\
  & = -\frac{3T\bar{\rho}\mu^2}{d} 
  \frac{\partial c^{\rss 2D}_{\rss sub}(K_1)}{\partial T} -
  \frac{T\bar{\rho}}{2d}\frac{\partial c^{\rss 2D}_{\rss sub}(0)}{\partial T}
  \eta^2 \nonumber \\
  & \approx   -\frac{3T\bar{\rho}\mu^2}{d} 
  \frac{\partial c^{\rss 2D}_{\rss sub}(K_1)}{\partial T} ,\label{entropy}
\end{align}
where we have used that $\delta \omega^{\rss 2D}_{\rss sub} =0$ along 
the melting line. The second term is of order $\eta^2\ll1$ and can be
neglected when compared to the first one. 
In order to calculate the entropy jump and $\eta_\mathrm{CC}$, we need to 
evaluate the partial derivative $\partial c^{\rss 2D}_{\rss sub}(K_1)/\partial T$ 
at melting.  The standard way \cite{ramakrishnan82,
cornaglia00} is to estimate $\partial c^{\rss 2D}_{\rss sub}(K_1)/\partial T$ 
from the temperature dependence of the solid structure factor. 
Here we proceed differently.
Comparing (\ref{etacomp}) with (\ref{cc1a}) (combined together 
with (\ref{entropy})), we see 
that in (\ref{etacomp}) the partial derivative of $c^{\rss 2D}_{\rss sub}$ 
with respect to $K_1$ appears whereas (\ref{cc1a}) contains the 
partial derivative with respect to $T$, once we use (\ref{entropy}) 
for the entropy jump. To compare the two different expressions 
for $\eta$ we need to find a way to connect these two partial 
derivatives. This relation can be found from the critical condition 
which determines the melting line as we show in the following.

The system freezes when the free energy at the solid minimum vanishes. 
Substituting  the values of $\mu_\mathrm{s}$ and 
of $\eta_\mathrm{s}$ at
the minimum into the expression for the free energy (\ref{freeen_ansatz}) (or, 
equivalently, into (\ref{freeen_constr})), 
the freezing  condition reads
\begin{displaymath}
   \frac{\delta \omega^{\rss 2D}_{\rss sub}(\eta_\mathrm{s},
   \mu_\mathrm{s})}{T} 
    = 3 c^{\rss 2D}_{\rss sub}(K_1) 
    \mu_\mathrm{s}^2 
     - (1+\eta_\mathrm{s})
     \Phi(c^{\rss 2D}_{\rss sub}(K_1)\mu_\mathrm{s}) = 0,
\end{displaymath}
where $K_1$ is related to $\eta_\mathrm{s}$ 
through (\ref{etak1}) and we have neglected in (\ref{freeen_ansatz}) 
the term  $(1-c^{\rss 2D}_{\rss sub}(0))\eta_\mathrm{s}^2$, 
which is quadratic in $\eta_\mathrm{s}$.
At the minimum, the molecular field $\xi_\mathrm{s}$
and the order parameter $\mu_\mathrm{s}$ are related through
$\xi_\mathrm{s} = c^{\rss 2D}_{\rss sub}(K_1) \mu_\mathrm{s}$, 
so the freezing condition can be rewritten as
\begin{equation}\label{min_xi_c}
  (3/c^{\rss 2D}_{\rss sub}(K_1))\xi_\mathrm{s}^2 
     - (1+\eta_\mathrm{s})
     \Phi(\xi_\mathrm{s}) = 0.
\end{equation}
For incompressible systems the same equation remains valid when one 
sets $\eta_\mathrm{s} = 0$
\begin{equation}\label{min_xi_in}
  (3/\bar{c}^{\rss 2D}_{\rss sub})\xi_\mathrm{s}^2 -
     \Phi(\xi_\mathrm{s}) = 0.
\end{equation}
From the discussion in the last section we know that
this equation is equivalent to the simple condition
$\bar{c}^{\rss 2D}_{\rss sub} = \bar{c}_c$, cf.\
(\ref{critical_inc}).  Comparing (\ref{min_xi_in})
with (\ref{min_xi_c}), it is easy to realize that
the freezing equation in the compressible theory is
equivalent to the one in the incompressible limit if
one replaces $c^{\rss 2D}_{\rss sub}(K_1)[1+\eta(K_1)]$
by $\bar{c}^{\rss 2D}_{\rss sub}$.  Therefore,
we can write a critical condition similar to
(\ref{critical_inc}) that is valid for a compressible
system,
\begin{equation}\label{critcomp}
  c^{\rss 2D}_{\rss sub}(T,K_1)
  [1 + \eta(K_1)] = \bar{c}_c,
\end{equation}
where we display explicitly both the $T$- and the $K_1$-dependences 
of the correlator. In (\ref{critcomp}), the correlator depends indirectly 
on the magnetic field $B$ in the solid phase through $K_1$ via
\begin{equation}\label{k1bm}
  K_1 = \Bigl( \frac{8\pi^2}{\sqrt{3}} \frac{B}{\Phi_0}\Bigr)^{1/2}
   \approx G(1+\eta/2).
\end{equation}
Consistently, since the magnetic field jumps across 
the transition ($\eta \neq 0$), the magnetic field in the 
liquid $B_\mathrm{l} = \Phi_0\bar\rho = B/(1+\eta)$ is different from 
the magnetic field $B$ in the solid phase.
Along the melting line $B_\mathrm{m}(T)$, $K_1$ can be written 
as a function of $T$ only by using (\ref{k1bm}), i.e. $K_1 = 
K_1(B_\mathrm{m}(T))$.
Hence, at melting, the LHS of (\ref{critcomp}) can 
be written as a function of the temperature alone; taking the 
total derivative $d/dT$ of (\ref{critcomp}) with respect to the temperature, 
we obtain
\begin{align}
\label{dini}
  \frac{\partial c^{\rss 2D}_{\rss sub}(T,K_1)}{\partial T}
  & =  - \frac{c^{\rss 2D}_{\rss sub}(T,K_1)}{1+\eta(K_1)} 
  \frac{\partial \eta(K_1)}{\partial K_1} 
   \frac{\partial K_1}{\partial B} \frac{d B_\mathrm{m}}{d T}
 \nonumber \\
  & 
  -\frac{\partial c^{\rss 2D}_{\rss sub}(T,K_1)}{\partial K_1} 
  \frac{\partial K_1}{\partial B} \frac{d B_\mathrm{m}}{d T}.
\end{align}
We need to compute the derivative
\begin{equation}\label{dhdg0}
  \frac{\partial K_1}{\partial B} 
  = \frac{K_1}{2B} \approx 
  \frac{G}{2B_\mathrm{l} (1+\eta/2)},
\end{equation} 
where we have used $B = B_\mathrm{l}(1+\eta)$ and 
the linearized relation between $K_1$ and $G$ in (\ref{k1bm}).
Inserting Eq.\ (\ref{dini}) (with the help of (\ref{dhdg0}) and 
(\ref{etak1})) into (\ref{entropy}), 
we obtain the entropy jump across the transition
\begin{align}
  \Delta s & = 
  \frac{3\mu^2 \bar{\rho} T}{d B_\mathrm{l}}
  \left[
  \frac{c^{\rss 2D}_{\rss sub}(T,K_1)}{1+\eta}\right.
   \nonumber \\ 
  &  \qquad + \left.\frac{G}{2(1+\eta/2)} 
  \frac{\partial c^{\rss 2D}_{\rss sub}(T,K)}{\partial K}
  \Bigr|_{K_1} 
  \right]\frac{d B_\mathrm{m}}{dT}.
\end{align}
\begin{figure}[t]
  \centering
   \includegraphics [width= 7.5cm] {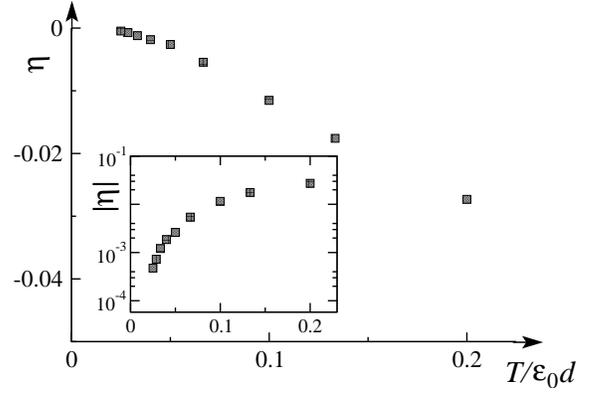}
   \caption
   [Jumps in density across the transition]
   {Values of the density jump $\eta$ across the transition
   as a function of temperature for the 3D vortex system. 
   The density jump $\eta$ is negative and small, of order 10$^{-4}$ at low
  $T$ (large magnetic fields) rising to 10$^{-2}$ at larger 
   $T$ (low magnetic fields). 
   Inset: log-plot of the absolute value of $\eta$.}
\label{fig:eta}
\end{figure}
Inserting this expression into (\ref{cc1a}), we obtain the density jump 
$\eta_\mathrm{CC}$
described by the Clausius-Clapeyron relation 
\begin{align}
    \eta_\mathrm{CC} 
   & = \frac{1}{1-c^{\rss 2D}_{\rss sub}(0)} \left[
    \frac{3c^{\rss 2D}_{\rss sub}(T, K_1)\mu^2}{1+\eta}\right. \nonumber \\
    & \qquad +\left.
    \frac{3G\mu^2}{2(1+\eta/2)}
  \frac{\partial c^{\rss 2D}_{\rss sub}(T,K)}{\partial K}\Bigr|_{K_1}
  \right],\label{cc1}
\end{align}
where in the last line we have used  
$d B_\mathrm{l}^2/4\pi \bar\rho T = d\Phi_0 
B_\mathrm{l} / 4\pi T= 1-c^{\rss 2D}_{\rss sub}(0)$ from (\ref{c3d0}). 
The first term in the square brackets in (\ref{cc1}) can be rewritten with the help
of the freezing condition. As a final result we obtain
\begin{align}
  \eta_\mathrm{CC} &  = \frac{1}{1-c^{\rss 2D}_{\rss sub}(0)}\Bigl[
  \Phi(c^{\rss 2D}_{\rss sub}(K_1) \mu) \nonumber \\
  & \quad +
    \frac{3G\mu^2}{2(1+\eta/2)}
  \frac{\partial c^{\rss 2D}_{\rss sub}(T,K)}{\partial K}\Bigr|_{K_1} \Bigr] = \eta, 
     \label{etacomp2}
\end{align}
which is exactly Eq.\ (\ref{etacomp}). Thus, we
conclude that the saddle point equation (\ref{etacomp})
is fully consistent with the Clausius-Clapeyron
relation.  

To obtain an estimate for the density jump across
the transition we have performed a numerical
minimization of the constrained free energy
(\ref{freeen_constr}). The system freezes
when the solid minimum exhibits the same free
energy as the liquid phase, i.e., when $\delta
\omega(\eta_\mathrm{s}, \mu_\mathrm{s}) =0$.  In Fig.\
\ref{fig:eta} we present the results of our numerical
analysis for various values of $T$.  For each $T$ we
show the density jump at the transition; as expected,
we find a negative value of $\eta$.  However, the
modulus of the density jump $\eta$ is always small,
between $|\eta| \approx 10^{-4}$ at low temperatures
(large $B$) and $|\eta| \approx 10^{-2}$ for large
temperatures (low $B$).  The effect of such a small
$\eta$ on the determination of the melting line and
on the value of $\mu_\mathrm{s}$ in the solid phase
is negligible. Finally, from (\ref{cc1a}) we can obtain the 
value of the entropy jump across the transition.  
In agreement with previous studies\cite{cornaglia00}, we obtain
a density jump per pancake vortex $\Delta S/N \approx 0.4\, 
k_\mathrm{B}$ at large fields decreasing 
to $\Delta S/N \approx 0.1\, k_\mathrm{B}$ at low magnetic fields
corresponding to temperatures of order 
$T = 0.2\, \varepsilon_0 d$.

\section{Conclusions}

We have presented an analysis of
the freezing transition of the magnetically coupled pancake vortex system
within a classical density functional theory. Despite the simplicity
of our approach, our results represent a considerable
improvement when compared to the predictions of earlier
work, particularly in the determination of the melting line. 
Moreover, we have addressed the problems with the thermodynamic 
inconsistency which affected earlier work. We showed how to 
obtain a negative density jump in accordance with the Clausius-Clapeyron
equation and the retrograded melting line. Note that our derivation of 
the Clausius-Clapeyron equation is not limited to the present 
vortex system but can be generalized to non-magnetic systems.

The techniques described in this paper
are easily extended to the study of inhomogeneous
situations, as was already done in the analysis of surface effects
on the melting transition\cite{decol05b}. 
The study of artificial surface pinning potentials offers itself as another
application of our DFT approach.
Such an analysis would shed light on the results of recent 
experiments described in Ref.\ \onlinecite{Fasano2005} 
where the response of vortices in BiSCCO to a weak
perturbation induced by pinning structures created on the sample
surface has been investigated. 

We acknowledge support from the Swiss National Foundation
through MaNEP (ADC) and from the DST (India) through
a Swarnajayanti Fellowship (GIM).


\begin{thebibliography}{31}
\bibitem{note_con}
Our approach neglects the effect of defects, such as interstitials 
and vacancies. If a finite density of defects was present, the 
perfect crystal constraint and the relation between $\eta$ and 
$K_1$ need to be modified.
\bibitem{note1}
In Ref.\ \onlinecite{chaikinlubensky} 
this expression is called the
Ursell function (up to a factor $1/\bar{\rho}^{\rss 3D}$).
\end{thebibliography}

\end{document}